\documentclass[12pt]{article}
\usepackage{amsmath,amsfonts}
\usepackage[dvips]{color}
\usepackage[nosort]{cite}
\usepackage{hyperref}
\usepackage{graphicx}
\usepackage{ulem}
\def\blfootnote{\xdef\@thefnmark{}\@footnotetext}

\long\def\symbolfootnote[#1]#2{\begingroup%
\def\thefootnote{\fnsymbol{footnote}}\footnote[#1]{#2}\endgroup}

\makeatletter
\def\@seccntformat#1{\csname the#1\endcsname.\quad}
\makeatother

\DeclareMathOperator{\diag}{diag}

\makeatletter\@addtoreset{equation}{section}\makeatother

\DeclareMathOperator{\Tr}{Tr}

\DeclareMathOperator{\sTr}{sTr}

\def\bC {\mathbb{C}}

\def\bP {\mathbb{P}}
\def\bR {\mathbb{R}}
\def\bZ {\mathbb{Z}}

\def\CP{\bC\bP}

\newcommand{\be}{\begin{equation}}
\newcommand{\ee}{\end{equation}}
\newcommand{\bea}{\begin{eqnarray}}
\newcommand{\eea}{\end{eqnarray}}

\newcommand{\vev}[1]{{\left< {#1} \right>}}

\newcommand{\eqn}[1]{(\ref{#1})}

\newcommand{\cA}{{\mathcal A}}

\newcommand{\cD}{{\mathcal D}}

\newcommand{\cN}{{\mathcal N}}
\newcommand{\cP}{{\mathcal P}}

\newcommand{\cR}{{\mathcal R}}

\renewcommand{\em}{\it}

\renewcommand{\title}[1]{\vbox{\center\LARGE
\bf\mathversion{bold}{#1}\mathversion{normal}
}\vspace{5mm}}
\renewcommand{\author}[1]{\vbox{\center#1}\vspace{5mm}}
\newcommand{\address}[1]{\vbox{\center\em #1}}
\newcommand{\email}[1]{\vbox{\center\tt#1}\vspace{5mm}}

\begin{document}

\begin{titlepage}
\begin{center}
\vspace{5mm}
\hfill {\tt HU-EP-09/30}\\
\hfill {\tt NSF-KITP-09-120}\\
\vspace{20mm} 
{\font\titlerm=cmr10 scaled\magstep4 {\titlerm 
A supermatrix model for 
\\[3mm]
{\LARGE $\cN=6$} super Chern-Simons-matter theory}}

\bigskip

\author{\large Nadav Drukker$^{1}$ and Diego Trancanelli$^{2,3}$}
\address{$^1$Institut f\"ur Physik, Humboldt-Universit\"at zu Berlin,\\
Newtonstra{\ss}e 15, D-12489 Berlin, Germany\\
\medskip
$^2$Department of Physics, University of California, \\
Santa Barbara, CA 93106, USA\\
\medskip
$^3$Kavli Institute for Theoretical Physics, University of California, \\
Santa Barbara, CA 93106, USA}

\email{drukker@physik.hu-berlin.de,\quad
dtrancan@physics.ucsb.edu}

\end{center}

\abstract{
\noindent
We construct the Wilson loop operator of $\cN=6$ super 
Chern-Simons-matter 
which is invariant under half of the supercharges of the theory and 
is dual to the simplest macroscopic open string in $AdS_4\times\CP^3$. 
The Wilson loop couples, in addition to the gauge and scalar fields of 
the theory, also to the fermions in the bi-fundamental representation 
of the $U(N)\times U(M)$ gauge group. These ingredients are naturally 
combined into a superconnection whose holonomy gives 
the Wilson loop, which can be defined for 
any representation of the supergroup $U(N|M)$. 
Explicit expressions for loops supported along an infinite straight  line and along 
a circle are presented. 
Using the localization calculation of Kapustin {\it et al.} we show that 
the circular loop is computed by a supermatrix model and discuss 
the connection to pure Chern-Simons theory with supergroup $U(N|M)$.
}

\vfill

\end{titlepage}


\section{Introduction}
\label{sec:intro}

The duality between string theory  on asymptotically $AdS$ spaces and 
conformal field theories has been an exciting area 
of research for over ten years now, with string theory providing 
answers to strong coupling questions in the gauge theory and vice-versa.

A year and a half ago, a new example of an $AdS$/CFT duality was
proposed by Aharony, Bergman, Jafferis, and Maldacena for the maximally 
supersymmetric gauge theory in three dimensions: $\cN=6$ supersymmetric 
Chern-Simons-matter with gauge group 
$U(N)\times U(N)$ \cite{ABJM}.%
\footnote{In this paper we will actually deal with  the generalization of this 
theory to the case of different ranks, $U(N)\times U(M)$, that was discussed 
in \cite{ABJ}.} 
The proposal was
inspired by a construction of the gauge theory with even more
supersymmetry, $\cN=8$, but which applied only to the gauge group
$SU(2)\times SU(2)$ \cite{Gustavsson:2007vu,Bagger:2007jr}.
The gravity dual of this theory is M-theory on $AdS_4\times S^7/\bZ_k$, where
$k$ is the level of the Chern-Simons term, or, for large enough $k$,
type IIA string theory on $AdS_4\times\CP^3$.

This gauge theory and the dual string theory have been studied extensively, 
but so far one of the most interesting observables in the gauge theory has not 
been constructed. Like in all gauge theories, one can define Wilson loop 
operators, which in the dual string theory are given by semi-classical 
string surfaces \cite{Maldacena-WL,Rey-Yee}. The most symmetric 
string of this type preserves half of the supercharges of the vacuum 
(as well as an $U(1)\times SL(2,\bR)\times SU(3)$ bosonic symmetry) but 
its dual operator in the field theory has not been identified yet.

So far the most symmetric Wilson loop operators in this theory, 
constructed in \cite{Dru-ABJM-wl, Chen-Wu-wl,Rey-ABJM-wl},  
preserve only $1/6$ of the supercharges 
and are therefore not viable candidates to be the dual of this classical string.
In fact, these operators exist also in Chern-Simons theories with
less supersymmetry \cite{Gaiotto:2007qi}
and do not get any supersymmetry enhancement
due to the clever quiver construction of the $\cN=6$ theory.

One reason to look for these operators is that Wilson loops are 
interesting observables in all gauge theories but in particular in Chern-Simons 
theories. In Chern-Simons without matter they are in fact the main observables. 
Beyond that, the lack of the gauge theory dual of the simplest string solution in 
$AdS_4\times\CP^3$ 
 is a glaring gap in our understanding of this duality.

As another motivation, recall that the analog observable in $\cN=4$ super 
Yang-Mills theory in four dimensions has the remarkable property that 
its expectation value is a non-trivial function of the coupling and of 
$N$ which can be calculated exactly by a Gaussian matrix model and 
interpolates from weak to strong coupling 
\cite{Erickson,Dru-Gross,Pestun:2007rz}. 

Another exact interpolating function which exists in the 4-dimensional theory is 
the cusp anomalous dimension, also known as the universal scaling function
\cite{Korchemsky:1988si,Gubser:2002tv,Frolov:2002av,Kruczenski:2002fb,BES,
Bern:2006ew,Benna:2006nd,Cachazo:2006az} 
which captures the scaling dimension of twist two operators. 
Trying to compute similar quantities in the 3-dimensional theory does not go 
through as nicely. In the calculation of the
spectrum of local operators there is a matching with the square root structure of 
the dispersion relation of giant magnons, but this involves one extra function of
the coupling \cite{Nishioka:2008gz,Gaiotto:2008cg,Grignani:2008is}, 
whose value is only known at weak and at strong coupling but not in 
the intermediate regime.

It is therefore interesting to revisit the question of Wilson loop operators 
in the hope that there are exact interpolating functions for them. 
For the $1/6$ BPS Wilson loop a matrix model has been recently derived 
in \cite{Kapustin:2009kz} and, despite its complexity, has been solved in 
the planar approximation in \cite{Marino-Putrov}.%
\footnote{This matrix model was studied also in \cite{Suyama:2009pd}.} 
Their results indeed match the string 
theory calculation and provide a first non-trivial interpolation function 
for this theory. 

Prompted by these considerations we construct here the 1/2 BPS Wilson 
loop for ${\cal N}=6$ super Chern-Simons-matter. Furthermore we prove that the 
results of \cite{Kapustin:2009kz, Marino-Putrov} carry over to our case. The calculation 
of \cite{Kapustin:2009kz} uses localization with respect to a specific 
supercharge which is also shared by the $1/2$ BPS loop. We show 
that the $1/2$ BPS Wilson loop is cohomologically equivalent to a very 
specific choice of the $1/6$ BPS loop and is therefore also given 
by a matrix model. This matrix model has a supergroup structure and 
the $1/2$ BPS loop is the most natural observable within this model. 
Indeed it can be calculated for all values of the coupling also beyond 
the planar approximation \cite{Marino-Putrov}.

In the coming section we present the loop and verify its symmetry. Our 
derivation uses in an essential way the quiver structure
of the theory. In addition to the gauge fields, the Wilson loop couples
to bilinears of the scalar fields and, crucially, also to the fermionic fields 
transforming in the bi-fundamental representation of the two gauge groups.
Our loop is classified by representations of the supergroup $U(N|M)$ and is 
defined in terms of the holonomy of a superconnection of this supergroup.%
\footnote{A somewhat similar construction for a topologically twisted 
version of ${\cal N}=4$ Chern-Simons-matter was shown in \cite{Kapustin:2009cd} 
to be equivalent to pure Chern-Simons theory with a supergroup.} 
In our analysis we consider both a loop supported along an infinite straight 
line and one supported along a circle.

In Section~\ref{sec:localization} 
we relate this Wilson loop to the $1/6$ BPS one and show that it 
is indeed the most natural observable for the matrix model of 
\cite{Kapustin:2009kz}. We interpret this matrix model as that of a supermatrix 
which represents the semiclassical expansion of pure Chern-Simons with 
supergroup $U(N|M)$ on the lens space $S^3/\mathbb{Z}_2$.

We conclude in Section 4 with a discussion of our results and some possible 
extensions. An appendix contains details about our notation.


\section{The loop}
\label{sec:loop}

We introduce now the construction of the Wilson loop in the 
$U(N)_k\times U(M)_{-k}$ Chern-Simons-matter theory.  
We denote the gauge field 
of the $U(N)$ factor as $A_\mu$ and the gauge field of the $U(M)$ factor 
as $\widehat A_\mu$. These gauge fields are coupled to four scalar 
fields $C_I$ and their complex conjugates $\bar C^I$, and to four fermions 
$\psi_{I}^\alpha$ and $\bar\psi^I_\alpha$, with $I=1,2,3,4$ being an $SU(4)_R$ index 
and $\alpha=+,-$ a spinor index. The scalars and the fermions are in 
the bi-fundamental representation of the gauge group. Our notation is 
such that $C\bar C$ and $\bar\psi\psi$ are in the adjoint of $U(N)$, 
whereas $\bar C C$ and $\psi\bar\psi$ are in the adjoint of $U(M)$. 
In the appendix we give more details about our 
conventions.

The central idea of this paper is to augment the connection of 
$U(N)\times U(M)$ to a superconnection of the form
\be
\label{sgroupac}
L\equiv
\begin{pmatrix}
A_\mu \dot x^\mu+\frac{2\pi}{k}|\dot x| M^I_JC_I\bar C^J  \quad
&
\sqrt{\frac{2\pi }{k}}\,|\dot x|\, \eta_I^\alpha \bar \psi^I_\alpha
\\
\sqrt{\frac{2\pi}{k}}\, |\dot x|\,\psi_I^\alpha \bar \eta^I_\alpha\quad
&
\widehat A_\mu \dot x^\mu +\frac{2\pi}{k}|\dot x|\widehat M^I_J \bar C^J C_I
\end{pmatrix},
\ee
where $x^\mu$ parametrizes the curve along which the loop operator is 
supported and $M^I_J$, $\hat M^I_J$, $\eta^\alpha_I$ and 
$\bar\eta^I_\alpha$ are free parameters.
A lot of the form of $L$ is dictated by dimensional analysis and by the index
structure of the fields. In three dimensions the scalars have dimension 
$1/2$, so they should appear as bi-linears, which are in the adjoint and therefore 
enter in the diagonal blocks
together with the gauge fields. The fermions have dimension $1$ and should
appear linearly. Since they transform in the bi-fundamental, they are naturally
placed in the off-diagonal entries of the matrix. 
Note that $\eta_I$ and $\bar\eta^I$ are Grassmann even, so that the
off-diagonal blocks of $L$ are Grassmann odd and $L$ is a
supermatrix.

Although $L$ has the structure of a $U(N|M)$ superconnection, 
the theory has only $U(N)\times U(M)$ gauge symmetry. It is nevertheless 
possible, given a path and the extra parameters, to calculate the holonomy 
of this superconnection and end up with a supermatrix. For a closed curve 
one can then take the trace%
\footnote{One could also take the supertrace instead of the trace. 
We will show later that supersymmetry imposes the latter.} in any representation $\cR$ of the supergroup 
$U(N|M)$. This gives the Wilson loop
\be
W_{\cal R}\equiv 
\Tr_{\cal R} {\cal P} \exp \left(i\int L \, d\tau\right)
\label{W}\,. 
\ee
%


\subsection{Infinite straight line}

In order to find the maximally supersymmetric Wilson loop, we
consider an operator defined along an infinite straight line in
the temporal direction, parameterized by $x^\mu=(\tau,0,0)$. 

The supercharges of this theory are
parameterized by the two-component spinors $\bar\theta^{IJ}_\alpha$ 
(see the appendix). 
Motivated by the $1/2$ BPS string solution in $AdS_4\times\CP^3$,
we want to find a loop operator
invariant under the same six supercharges. They are in fact the same
supercharges also annihilated by other brane solutions dual to the vortex
loop operators of \cite{Dru-vortex} and are parameterized by
\be \bar\theta^{1I}_+\,,\qquad \bar\theta^{IJ+}\,,\qquad
I,J=2,3,4\,. \label{supercharges}\ee

As mentioned before, this loop should also preserve an $SU(3)$ subgroup
of the $R$-symmetry group. Given that and the chirality of the supercharges, 
this suggests the ansatz
\be
M^I_J= \widehat M^I_J=m_1 \,\delta^I_J-2m_2\,\delta^I_1\delta_J^1\,,
\qquad
\eta_I^\alpha=\eta\,\delta_I^1\delta^\alpha_+\,,
\qquad
\bar\eta^I_\alpha=\bar\eta\,\delta^I_1\delta_\alpha^+\,.
\label{MIJ}
\ee

We define the modified connections which appear in the diagonal blocks 
of $L$
\be \cA_0\equiv A_0+\frac{2\pi}{k} M^I_JC_I\bar C^J\,, \qquad
 \widehat\cA_0\equiv  \widehat A_0+\frac{2\pi}{k} \widehat M_J^I\bar C^JC_I\,.
\label{mod-A}
\ee
One can easily verify \cite{Dru-ABJM-wl, Chen-Wu-wl,Rey-ABJM-wl} 
that the supersymmetry variation of these terms does not vanish. 
Instead we demand that their variation 
contains only $\psi_1^+$ and $\bar\psi_+^1$, which appear anyhow in the 
Wilson loop through the couplings to $\eta^\alpha_I$ and $\bar\eta_\alpha^I$. 
Using the expressions in the appendix we find that for the particular choice%
\footnote{This value of $M^I_J$ is the same as the effective mass matrix of
W-bosons arising upon higgsing the gauge symmetry
\cite{Berenstein:2008dc} (see also \cite{Rey-ABJM-wl}).}
of $m_1=m_2=1$ the 
variation is
(noticing that $\psi^+=\psi_-$ and $\psi_+=-\psi^-$)
\be
\begin{aligned}
 \delta\cA_0&=\frac{8\pi}{k}\left[\bar\theta^{1I}_+\,C_I\,\psi_{1}^+
-\frac{1}{2}\varepsilon_{1IJK}\,\bar\theta^{IJ+}\,\bar\psi^1_+\,\bar
C^K\right]\,,\\
\delta \widehat\cA_0&=\frac{8\pi}{k}\left[\bar\theta^{1I}_+\, \psi_{1}^+\, C_I
-\frac{1}{2}\varepsilon_{1IJK}\,\bar\theta^{IJ+}\,\bar C^K
\,\bar\psi^1_+ \right]\,.
\label{var-bosons}
\end{aligned}
\ee

Turning to the off-diagonal entries in $L$, the variation of the 
fermions $\bar\psi^1$ and $\psi_1$ includes the covariant derivative 
$\gamma^\mu D_\mu$. Since the fermions appearing in the loop have 
specific chiralities, as do the supercharges \eqn{supercharges}, 
the covariant derivative gets projected to be along the direction of the loop 
by
\be
(i\gamma^\mu)_+^{\;\;+}=\delta_0^\mu\,, \qquad
(i\gamma^\mu)_-^{\;\;-}=-\delta_0^\mu\,.
\ee
Furthermore, all the non-linear terms appearing in the variation of the 
fermions can be repackaged into a covariant derivative with the modified 
connection \eqn{mod-A}
\be
\begin{aligned}
\cD_0 C_I &= \partial_0 C_I + i (\cA_0\, C_I - C_I\, \widehat \cA_0)\,, \\
\cD_0 {\bar C}^{I} &= \partial_0 {\bar C}^{I} 
- i ({\bar C}^{I}\,\cA_0 -   \widehat\cA_0\, {\bar C}^{I}) \,,
\end{aligned}
\label{cov-deriv}
\ee
with exactly the choice (\ref{MIJ}) of $M^I_J$ and $ \widehat M^I_J$.

We finally find that
\be
\begin{aligned}
\delta\bar\psi^1_+&
=2\bar\theta^{1I}_+{\cal D}_0 C_I\,, \\\
\delta\psi^+_1&=
-\varepsilon_{1IJK}\bar\theta^{IJ+}{\cal D}_0 \bar C^K\,.
\label{var-fermions}
\end{aligned}
\ee
Combining \eqn{var-bosons} and \eqn{var-fermions} the variation of $L$ 
for the time-like line is given by
\be
\begin{aligned}
\delta L&=
\frac{8\pi}{k}\bar\theta^{1I}_+
\begin{pmatrix}
C_I\,\psi_{1}^+\quad&
\sqrt{\frac{k}{8\pi}}\eta\cD_0C_I\\
0&
\psi_{1}^+\, C_I
\end{pmatrix}
-\frac{4\pi}{k}\varepsilon_{1IJK}\,\bar\theta^{IJ+}
\begin{pmatrix}
\bar\psi^1_+\,\bar C^K&
0\\
\sqrt{\frac{k}{8\pi}}\bar\eta\cD_0\bar C^K\quad&
\bar C^K\,\bar\psi^1_+
\end{pmatrix}\,.
\end{aligned}
\label{var-L}
\ee

The proof of supersymmetry-invariance of the Wilson loop requires one 
additional step, namely integration by parts.
Expanding to second order, the Wilson loop is
\be
W_{\cal R}
=\Tr_{\cal R}
\left[1+i\int_{-\infty}^\infty d\tau\, L(\tau)-\int_{-\infty}^\infty
d\tau_1 \int_{\tau_1}^{\infty}d\tau_2\,
L(\tau_1)L(\tau_2)+\ldots\right]\,.
\label{expansionW}
\ee
The off-diagonal pieces of the linear term are total derivatives, as can be seen 
in \eqn{var-L} and integrate away. 
The diagonal part of the linear term does not vanish on its own, but it is
canceled by the variation of the fermions in the quadratic term.
To see that, we write 
the relevant terms for the variations with parameters $\bar\theta^{1I}_+$
\be
\begin{aligned}
\delta W_{\cal R}
&=\frac{8\pi}{k}\bar\theta^{1I}_+
\Tr_{\cal R}
\Bigg[i\int_{-\infty}^\infty d\tau\,
\begin{pmatrix}
C_I\,\psi_{1}^+\quad&
\\
&
\psi_{1}^+\, C_I
\end{pmatrix}
\\&\hskip1.5cm
-\frac{1}{2}\eta\bar\eta\int_{-\infty}^\infty
d\tau_1 \int_{\tau_1}^{\infty}d\tau_2\,
\begin{pmatrix}
\partial_{\tau_1}C_I(\tau_1)\psi_1^+(\tau_2)&\\
 &-\psi_1^+(\tau_1)\partial_{\tau_2}C_I(\tau_2)
\end{pmatrix}
+\ldots\Bigg].
\end{aligned}
\label{delta-W}
\ee
The last entry on the bottom right comes from the variation of $L(\tau_2)$ and
it has an extra minus sign since the supersymmetry parameter $\bar\theta$ was
permuted through the first fermion $\psi_1^+(\tau_1)$. We have also assumed that
$\eta$ and $\bar\eta$ are constant, so we have pulled them out of the integrals.

Integrating by parts and ignoring any possible contributions from infinity, one obtains
\be
\begin{aligned}
\frac{8\pi}{k}\bar\theta^{1I}_+
\Bigg[i\int_{-\infty}^\infty d\tau\,
\begin{pmatrix}
C_I\,\psi_{1}^+&\\
&
\psi_{1}^+\, C_I
\end{pmatrix}
-\frac{1}{2}\eta\bar\eta\int_{-\infty}^\infty d\tau
\begin{pmatrix}
C_I\psi_1^+
&\\
&\psi_1^+C_I
\end{pmatrix}
\Bigg]\,.
\end{aligned}
\ee
The two integrals clearly cancel each other for $\eta\bar\eta=2i$. A similar
cancellation takes place for the $\bar\theta^{IJ+}$ supercharges.

To summarize, we have shown at leading order in the expansion
(\ref{expansionW}) that the Wilson loop
(\ref{sgroupac}), (\ref{W}) with 
\be M^I_J= \widehat M^I_J= \delta^I_J-2\delta^I_1\delta_J^1\,,\qquad
\eta_I^\alpha=\eta\,\delta_I^1\delta^\alpha_+\,,
\qquad
\bar\eta^I_\alpha=\bar\eta\,\delta^I_1\delta_\alpha^+\,,
\qquad
\eta\bar\eta=2i\,,\ee
 preserves
the six Poincar\'e supercharges (\ref{supercharges}) and is
therefore $1/2$ BPS.
We performed the same calculation to the next loop order by
multiplying the diagonal part of $\delta L$ with another $L$ and
the off-diagonal pieces with two more and integrating one of the
three integral by parts. After including all the terms, the final result
vanishes again.

This analysis can be carried over to all orders. To do that we separate
$L$ into the diagonal part $L_B$ and the off-diagonal entries $L_F$.
We leave the bosonic piece in the exponent and expand only $L_F$
\be
W_{\cal R}
=\Tr_{\cal R}\cP\left[
e^{i\int L_B \, d\tau}
\left(
1+i\int_{-\infty}^\infty d\tau_1\, L_F(\tau_1)-\int_{-\infty}^\infty
d\tau_1 \int_{\tau_1}^{\infty}d\tau_2\,
L_F(\tau_1)L_F(\tau_2)+\ldots\right)\right].
\label{LBLF}
\ee
The supersymmetry variation can act on the exponent, bringing down
an extra integral of $\delta L_B$ or can act on one of the $L_F$, giving
a matrix with an off-diagonal entry of the form $\cD_0 C_I$ (or $\cD_0\bar C^K$).
As mentioned before, this $\cD_0$ is the covariant derivative with
the modified connection appearing in $L_B$. This allows us to integrate
by parts these terms in the presence of the path ordered $\exp(i\int L_B)$.
As in the case considered explicitly above, these will give non-zero
contributions at the limits of integration, where in general we have
\begin{align}
\label{allorder}
&
i^p\int_{\tau_1<\cdots<\tau_p}\hskip -1cm d\tau_1\cdots d\tau_n\cdots d\tau_p\,
L_F(\tau_1)\cdots\delta L_F(\tau_n)\cdots L_F(\tau_p)
\nonumber\\&\hskip.5cm
\propto
(-1)^{n-1}i^p\, \bar\theta^{1I}_+
\int_{\tau_1<\cdots<\text{\sout{$\tau_n$}}<\cdots<\tau_{p}}\hskip -2cm d\tau_1 \cdots
\,\text{\sout{$d\tau_n$}}\,\cdots d\tau_p
\\&\hskip1cm
L_F(\tau_1)\cdots
\left[L_F(\tau_{n-1})
\begin{pmatrix}
(C_I\psi_1^+)(\tau_{n+1})&0\\0&0
\end{pmatrix}
+\begin{pmatrix}
0&0\\0&(C_I\psi_1^+)(\tau_{n-1})
\end{pmatrix}
L_F(\tau_{n+1})\right]
\cdots L_F(\tau_p)\,,
\nonumber
\end{align}
with the factor $(-1)^{n-1}$ coming from pulling out $\bar\theta^{1I}_+$ 
through the $L_F$ insertions. 
Reordering the terms we see that this exactly cancels the insertion of the
variation $\int \delta L_B$ into the term in \eqn{LBLF} with $p-2$ integrals
of $L_F$.

This calculation proves that this Wilson loop preserves six of the twelve Poincar\'e 
supercharges. Similarly, one can show that six conformal supercharges are also
preserved.


\subsection{Circle}

Under a conformal transformation a line is transformed into a
circle. While conformal transformations are a symmetry of the
theory, they change the topology of the curve and, as it turns out,
also the expectation value of the loop. In the case of the 1/2 BPS
Wilson loops of $\cN=4$ super Yang-Mills one finds that, whereas the
straight line has trivial expectation value, the circular loop
depends in an interesting way on the coupling constant of the
theory. It is therefore of great interest to consider circular
Wilson loops also in this 3-dimensional theory.

First we consider the Wick rotation of the time-like line to a space-like line.
The latter can be defined either for the theory in Euclidean $\bR^3$ or
in the Lorentzian theory in $\bR^{1,2}$, as we do here. 
Indeed it is simple to check that the replacement 
$|\dot x|\to -i|\dot x|$ gives the $1/2$ BPS Wilson loop for a space-like line. 
This replacement affects the scalar bi-linear term and the fermionic terms.

To get the circle one should perform a conformal transformation. The path 
is now given by%
\footnote{We consider for simplicity a circle of unit radius but the 
result is invariant under conformal transformations and applies to 
an arbitrary radius circle.}
\be
x^1=\cos\tau\,,\qquad
x^2=\sin\tau\,.
\ee
The scalar couplings should not be affected by the conformal transformation, 
so for the diagonal 
part of the superconnection $L$ \eqn{sgroupac} we again use the shorthands
\be
\cA\equiv A_\mu\dot x^\mu-i\frac{2\pi}{k}
M^I_JC_I\bar C^J\,,\qquad
\widehat\cA\equiv  \widehat A_\mu\dot x^\mu-i\frac{2\pi}{k}
\widehat M^I_J\bar C^JC_I\,.
\label{cA-circle}
\ee

We still should couple only to the fermion fields $\psi_1^\alpha$ and $\bar\psi^1_\alpha$. 
The spinor index is chosen by taking $\eta^\alpha_I(\tau)$ and $\bar\eta_\alpha^I(\tau)$ 
to be eigenstates of the projector
\be
1+\dot x^\mu(\gamma_\mu)_\alpha^{\ \beta}
=\begin{pmatrix}
1&-ie^{-i\tau}\\ie^{i\tau}&1
\end{pmatrix},
\label{projectoreigenstates}
\ee
thus
\be
\eta_{I}^\alpha(\tau)
=
\begin{pmatrix}
1\ &-ie^{-i\tau}
\end{pmatrix}\eta(\tau)\,\delta^1_I\,,
\qquad
\bar\eta_{\alpha}^I(\tau)
=i\begin{pmatrix}
1\\i e^{i\tau}
\end{pmatrix}\bar \eta(\tau)\,\delta^I_1\,,
\label{couplings-circle-0}
\ee
with an arbitrary function $\eta(\tau)$ which is determined by checking the 
supersymmetry variation of the loop.
 
The loop along the line preserved six super-Poincar\'e symmetries and six 
superconformal ones, for the circle we expect to find twelve which are linear 
combinations of the two. The parameters of the superconformal 
transformations, which we label $\bar\vartheta^{IJ}$, should be related 
to the super-Poincar\'e transformation parametrized by $\bar\theta^{IJ\alpha}$. 
We take the ansatz
\be
\bar\vartheta^{1I\alpha}=i\,\bar\theta^{1I\beta}(\sigma^3)_\beta^{\ \alpha}\,,
\qquad
\bar\vartheta^{IJ\alpha}=-i\,\bar\theta^{IJ\beta}(\sigma^3)_\beta^{\ \alpha}\,,
\qquad
I,J\neq1\,,
\label{vartheta}
\ee
and using the explicit superconformal transformations \cite{Bandres:2008ry} 
determine the supersymmetric circular loop. Note that the choice in 
\eqn{vartheta} is consistent with the reality condition (\ref{real}) on 
$\bar\theta^{IJ}$ and the analog one for $\bar\vartheta^{IJ}$.

To do the calculation we note that, apart for one extra term in the variation
of the spinors, the superconformal transformations of the fields are the same as the super-Poincar\'e transformations,
modulo the replacement $\bar\theta^{IJ}\to\bar\vartheta^{IJ}x^\mu\gamma_\mu$.
Using that
$x^\mu\gamma_\mu\dot x^\nu\gamma_\nu=i\sigma^3$,
with our choice of $\vartheta^{IJ}$ we find
\be
\begin{aligned}
\bar\theta^{1I}+\bar\vartheta^{1I}x^\mu\gamma_\mu
&=\bar\theta^{1I}(1-\dot x^\mu\gamma_\mu)\,,
\\
\bar\theta^{IJ}+\bar\vartheta^{IJ}x^\mu\gamma_\mu
&=\bar\theta^{IJ}(
1+\dot x^\mu\gamma_\mu)\,,\qquad
I,J\neq1\,.
\end{aligned}
\label{P+S-charges}
\ee
Another useful relation involves the change of spinor indices on the
projector
\be
(1\pm\dot x^\mu\gamma_\mu)_\alpha{}^\beta
=(-1\pm\dot x^\mu\gamma_\mu)^\beta{}_\alpha\,.
\ee
As mentioned in the appendix, unless we write it explicitly, we always use the indices as
in the left-hand side of this equation. Lastly, we note that
\be
(1\pm\dot x^\nu\gamma_\nu)\gamma^\mu(1\pm\dot x^\rho\gamma_\rho)
=\pm2(1\pm\dot x^\nu\gamma_\nu)\dot x^\mu\,.
\ee
Using these relations we get the variations under Poincar\'e and 
superconformal transformations of the fields in $L$
\be
\begin{aligned}
\delta\cA
&=\frac{8\pi i}{k}\bar\theta^{1I}(1-\dot x^\mu\gamma_\mu)C_I\psi_1
+\frac{4\pi i}{k}\varepsilon_{1IJK}\bar\theta^{IJ}
(1+\dot x^\mu\gamma_\mu)\bar\psi^1\bar C^K\,,
\\
\delta \widehat\cA
&=\frac{8\pi i}{k}\bar\theta^{1I}(1-\dot x^\mu\gamma_\mu)\psi_1C_I
+\frac{4\pi i}{k}\varepsilon_{1IJK}\bar\theta^{IJ}
(1+\dot x^\mu\gamma_\mu)\bar C^K\bar\psi^1\,,
\\
\delta(\eta_1^\alpha(\tau)\bar\psi^1_\alpha)
&=4i\eta_1\bar\theta^{1I}\dot x^\mu\cD_\mu C_I
-2\eta_1\sigma^3\bar\theta^{1I}C_I\,,
\\
\delta(\psi_1^\alpha\bar\eta^1(\tau)_\alpha)
&=-\varepsilon_{1IJK}\bar\theta^{IJ}
\Big[2i\bar\eta^1\dot x^\mu\cD_\mu \bar C^K
+\sigma^3\bar\eta^1\bar C^K)\Big]\,.
\end{aligned}
\ee
The extra terms in the variations of $\psi$ and $\bar\psi$ 
are written explicitly in \cite{Dru-vortex}. We would 
like to write the last two expressions as total derivatives, which gives the equations 
\be
\partial_\tau\eta_1=\frac{i}{2}\eta_1\sigma^3\,,\qquad
\partial_\tau\bar\eta^1=-\frac{i}{2}\sigma^3\bar\eta^1\,.
\ee
From this we deduce that the extra function $\eta(\tau)$ in \eqn{couplings-circle-0} 
is
$ \eta(\tau)=e^{i\tau/2}$. The product of the two couplings is then 
$\eta^\alpha_I\bar\eta^I_{\alpha}=2i$, as in the case of the 
line.

The superconnection for the circular Wilson loop is therefore
\be
\label{L-circle}
L\equiv
\begin{pmatrix}
\cA\quad
&
-i\sqrt{\frac{2\pi }{k}}\, \eta_I^\alpha \bar \psi^I_\alpha
\\
-i\sqrt{\frac{2\pi}{k}}\, \psi_I^\alpha \bar \eta^I_\alpha\quad
&
\widehat \cA
\end{pmatrix},
\ee
with $\cA$ and $\widehat\cA$ defined in \eqn{cA-circle} and
\be
\eta_{I}^\alpha(\tau)
=
\begin{pmatrix}
e^{i\tau/2}\ &-ie^{-i\tau/2}
\end{pmatrix}\delta^1_I\,,
\qquad
\bar\eta_{\alpha}^I(\tau)
=\begin{pmatrix}
ie^{-i\tau/2}\\- e^{i\tau/2}
\end{pmatrix}\delta^I_1\,,
\label{couplings-circle}
\ee

Collecting all the pieces, we find that the variation is
\be
\begin{aligned}
\delta L&=
\frac{8\pi i}{k}
\begin{pmatrix}
C_I\,\psi_{1}(1+\dot x^\mu\gamma_\mu)\quad&
-i\sqrt{\frac{k}{2\pi}}\cD_\tau(\eta_1C_I)\\
0&
\psi_{1}\, C_I(1+\dot x^\mu\gamma_\mu)
\end{pmatrix}
\bar\theta^{1I}
\\&\qquad
+\frac{4\pi i}{k}\varepsilon_{1IJK}\,\bar\theta^{IJ}
\begin{pmatrix}
(1+\dot x^\mu\gamma_\mu)\bar\psi^1\,\bar C^K\quad&
0\\
i\sqrt{\frac{k}{2\pi}}\cD_\tau(\bar\eta^1\bar C^K)\quad&
(1+\dot x^\mu\gamma_\mu)\bar C^K\,\bar\psi^1
\end{pmatrix}\,.
\end{aligned}
\label{var-L-circle}
\ee

It is instructive to repeat the supersymmetry analysis at leading order also for the circle. Expanding the loop as in (\ref{expansionW}) and varying it as in (\ref{var-L-circle}), one finds (we consider just one kind of supercharges and write explicitly only the terms in the diagonal blocks)
\begin{align}
\delta W_{\cal R}&\propto 
i\Tr_{\cal R}\int_0^{2\pi} d\tau
\begin{pmatrix}
C_I\,\psi_{1}(1+\dot x^\mu\gamma_\mu)\bar\theta^{1I}
&
\\
&
\psi_{1}\, C_I(1+\dot x^\mu\gamma_\mu)\bar\theta^{1I}
\end{pmatrix}\\
&\hskip 1cm -\Tr_{\cal R}
\int_0^{2\pi}
d\tau_1\int_{\tau_1}^{2\pi} d\tau_2
\begin{pmatrix}
-(\partial_{\tau_1}\eta_1 C_I \bar\theta^{1I})_{(1)}(\psi_1 \bar\eta^1)_{(2)}
&
\\
&
-(\psi_1 \bar\eta^1)_{(1)}(\partial_{\tau_2}\eta_1 C_I \bar\theta^{1I})_{(2)}
\end{pmatrix}\,.
\nonumber
\end{align}
As done for the line, it is easy to integrate by parts and verify the cancellation of the bulk terms between the first and the second lines of this expression. From the integration by parts one has now also the following boundary terms
\be
-\Tr_{\cal R}\int_0^{2\pi} d\tau\begin{pmatrix}
\left(\eta_1 C_I \bar\theta^{1I}\right)\!(0) \,\,(\psi_1\bar\eta^1)(\tau)&
\\
&- (\psi_1\bar\eta^1)(\tau)\, \left(\eta_1 C_I \bar\theta^{1I} \right)\!(2\pi)
\end{pmatrix}\,,
\label{boundarycircle}
\ee
which cancel once taking the trace, since $\eta_1$ is antiperiodic on the circle, 
$\eta_1(2\pi)=-\eta_1(0)$. This calculation in fact determines that the Wilson 
loop is supersymmetric only when taking the trace of the holonomy, 
and not the supertrace.%
\footnote{The trace of a supermatrix in an arbitrary representation 
is defined on diagonal matrices by the supertrace as 
$\Tr_{\cR}\big(\begin{smallmatrix}a&0\\0&b\end{smallmatrix}\big)
=\sTr_{\cR}\big(\begin{smallmatrix}a&0\\0&-b\end{smallmatrix}\big)$. 
With this definition, equation \eqn{boundarycircle} vanishes for any representation.
\label{tr-footnote}}

We can repeat the all-order proof outlined in (\ref{allorder}).
Expanding the exponential in $L_F$
one can see again cancellations between bosons and fermions similarly to
what happened for the line. The only difference from that case 
are the new boundary terms arising at $\tau=0$ from integrating over the first 
variation $\delta L_F(\tau_1)$ and at $\tau=2\pi$ from the last variation 
$\delta L_F(\tau_p)$. As in the leading order case studied above, 
upon taking the trace these two contributions cancel. 

The same analysis carried out above 
applies also to the six other 
supercharges. We have shown then that the circle operator 
is invariant under the twelve supercharges in 
(\ref{P+S-charges}) and is therefore 1/2 BPS.


\section{Localization to a matrix model}
\label{sec:localization}

Recently, in a very nice paper \cite{Kapustin:2009kz}, the evaluation of 
supersymmetric Wilson loop operators in Chern-Simons-matter theories 
with $\cN=2$ supersymmetry was reduced to a 0-dimensional matrix model. 
In this section we show how to apply the same result to 
the circular Wilson loop constructed in the preceding section.%
\footnote{The calculation of \cite{Kapustin:2009kz} was done 
for a Wilson loop on the equator of $S^3$, while here we discuss a circle 
in $\bR^3$ or $\bR^{1,2}$. These operators should have the same 
expectation value when normalizing by the partition function.} 
 We can then use the solution of this matrix model 
\cite{Marino-Putrov} to evaluate the Wilson loop at arbitrary values of the 
coupling constants.

The Wilson loop studied in \cite{Kapustin:2009kz} is the one constructed in 
\cite{Gaiotto:2007qi}. The $\cN=2$ Chern-Simons-matter theories have 
supersymmetric Wilson loops with the gauge connection 
and an extra coupling to the scalar in the vector multiplet. This scalar has an 
algebraic equation of motion and after integrating it out we find the Wilson loop 
with a coupling to some of the other scalar fields of the theory.

Specializing to the case of the theory with $\cN=6$ supersymmetry, one 
ends up with the Wilson loops of the type constructed in 
\cite{Dru-ABJM-wl, Chen-Wu-wl,Rey-ABJM-wl}, where the connection is 
given by \eqn{sgroupac} with $\eta_I^\alpha=\bar\eta^I_\alpha=0$ and 
$M^I_J=\widehat M^I_J=\diag(-1,-1,1,1)$.%
\footnote{The choice of $1/6$ BPS Wilson loop is not unique. The constructions 
in \cite{Dru-ABJM-wl, Chen-Wu-wl,Rey-ABJM-wl} all preserve the same 
supercharges but have slight differences. We use this definition, since 
it will turn out to be related to the loop constructed in Section~\ref{sec:loop}, 
as we show below.}
In the following we will denote the resulting connection matrix by $L_{1/6}$ and 
that for the loop constructed in Section~\ref{sec:loop} by $L_{1/2}$.
The reason for this notation is that while these Wilson loops preserve half of the 
supercharges of the the $\cN=2$ theories, they do not see 
the supersymmetry enhancement of the gauge theory from $\cN=2$ to $\cN=6$, 
so they are $1/6$ BPS. The loops 
constructed in Section~\ref{sec:loop} preserve instead  half of the supercharges of
the $\cN=6$ theory.


\subsection{Relation between the different Wilson loops}

The calculation of \cite{Kapustin:2009kz} uses localization with respect to 
a single supercharge, which is also shared by the $1/2$ BPS Wilson loop. 
We will show now that the $1/2$ BPS Wilson loop is related to the 
$1/6$ BPS loop --- they are in the same cohomology class under this 
supercharge. Hence the localization calculation immediately applies also 
to the $1/2$ BPS Wilson loop.

We start by analyzing the case of the infinite straight  line. We notice
that the $1/2$ BPS Wilson loop shares all four supercharges preserved by 
the $1/6$ BPS one. These are the ones parameterized by 
$\bar\theta^{12}_+$ and $\bar\theta^{34+}$ and their superconformal counterparts.
For the $1/6$ BPS Wilson loop the couplings of the scalars is given by the 
matrices $M^I_J=\widehat M^I_J=\diag(-1,-1,1,1)$ and there is no coupling to the fermions. 
One can therefore write the difference between the superconnection for the $1/2$ BPS loop and the connection of the $1/6$ BPS one as
\be
\tilde L = L_{1/2}-L_{1/6}=
\begin{pmatrix}
\frac{4\pi}{k}C_2\bar C^2  \quad
&
\sqrt{\frac{2\pi }{k}}\,\eta\, \bar \psi^1_+
\\
\sqrt{\frac{2\pi}{k}}\,\bar\eta\,\psi_1^+\quad
&
\frac{4\pi}{k}\bar C^2 C_2
\end{pmatrix}\, .
\ee
The off-diagonal term in $\tilde L$ is the same as $L_F$ defined above in \eqn{LBLF}. 
The diagonal piece comes from the difference in the scalar couplings 
$M_J^I$ and $\widehat M_J^I$ between the two loops.

We want now to show that the 1/2 BPS loop, $W_{1/2}$, and the 1/6 BPS one, 
$W_{1/6}$, are cohomologically equivalent with respect to the aforementioned 
supercharges. This means that the difference between the two loops is exact 
with respect to a linear combination $Q$ of the supersymmetries with parameters 
$\bar\theta^{12}_+$ and $\bar\theta^{34+}$, namely that there exists a $V$ such 
that
\be
W_{1/2}-W_{1/6} = \Tr_{\cal R} {\cal P} \left[e^{i\int L_{1/2}}-e^{i\int L_{1/6}}\right] = Q\, V\,,
\qquad Q\equiv Q_{12}^+ + Q_{34+}\,.
\ee
To find $V$  it is useful to rewrite the difference between the loops as
\be
W_{1/2}-W_{1/6}
=\Tr_{\cal R}\cP\left[
e^{i\int_{-\infty}^\infty L_{1/6}(\tau)d\tau} \sum_{p=1}^\infty\,  i^p
\int_{-\infty<\tau_1<\cdots<\tau_p<\infty}\hskip-2cm d\tau_1\cdots d\tau_p\,
\tilde L(\tau_1)\cdots\tilde L(\tau_p)\right]\,.
\label{W-difference}
\ee
We take
\be
V=i \Tr_{\cal R}\cP\left[\int_{-\infty}^\infty d\tau \,
e^{i\int_{-\infty}^\tau L_{1/6}(\tau_1)d\tau_1}
\,\Lambda(\tau)\,
e^{i\int_\tau^\infty L_{1/2}(\tau_2)d\tau_2}
\right]\,,
\ee
where 
\be
\Lambda= \sqrt{\frac{\pi }{2k}}\,
\begin{pmatrix}
0
&
-\eta\, C_2
\\
\bar\eta\, \bar C^2
&
0
\end{pmatrix}
\ee
is such that $Q\, \Lambda = L_F$.
Acting with $Q$ on $V$ and recalling that $Q\, L_{1/6}=0$, one finds the following two terms
\be
Q\, V=i \Tr_{\cal R}\cP\left[\int_{-\infty}^\infty d\tau \,
e^{i\int^\tau_{-\infty} L_{1/6}(\tau_1)d\tau_1}\left(
L_F(\tau) e^{i\int_{\tau}^{\infty} L_{1/2}(\tau_2)d\tau_2}
+ \Lambda(\tau)\, 
Q\, e^{i\int_{\tau}^{\infty} L_{1/2}(\tau_2)d\tau_2}
\right)
\right]\,.
\label{QV}
\ee
To evaluate the second contribution we can use a similar logic to the all-order proof (\ref{allorder}) and Taylor expand the $L_F$ in the exponent, the difference being that the integral in the exponent is now between $\tau$ and infinity rather than between minus infinity and infinity. The cancellation between bosons and fermions is therefore incomplete and when $Q$ acts on the first $L_F$ the integration by parts introduces an extra boundary term
\be
i \Lambda(\tau) \int_\tau^{\infty}d\tau_1\, Q\, L_F(\tau_1)=
\begin{pmatrix}
\frac{4\pi}{k} (C_2\bar C^2)(\tau)  \quad
&
0
\\
0
&
\frac{4\pi}{k}(\bar C^2 C_2)(\tau)
\end{pmatrix}\,.
\label{QVboundterm}
\ee
This is nothing else than the diagonal part of $\tilde L$. Combining it with the term in $L_F$ in (\ref{QV}), one finds
\bea
Q\, V& =&i \Tr_{\cal R}\cP\left[\int_{-\infty}^\infty d\tau \,
e^{i\int_{-\infty}^\tau L_{1/6}(\tau_1)d\tau_1}
\tilde L(\tau)\,
e^{i\int_\tau^\infty L_{1/2}(\tau_2)d\tau_2}
\right]\cr
& =&i \Tr_{\cal R}\cP\left[
e^{i\int_{-\infty}^\infty L_{1/6}(\tau_1)d\tau_1}
\int_{-\infty}^\infty d\tau \,
\tilde L(\tau)\,
e^{i\int_\tau^\infty \tilde L(\tau_2)d\tau_2}
\right]\,,
\eea
which, upon Taylor expansion, can be seen to be exactly equal to (\ref{W-difference}).

We analyze now the circular loop. The difference between the connections is now
\be
\tilde L=L_{1/2}-L_{1/6}=
\begin{pmatrix}
-i\frac{4\pi}{k}C_2\bar C^2  \quad
&
-i\sqrt{\frac{2\pi }{k}}
\,\eta^\alpha_1(\tau)\, \bar \psi^1_\alpha
\\
-i\sqrt{\frac{2\pi}{k}}
\,\psi_1^\alpha\,  \bar\eta^1_\alpha(\tau)\quad
&
-i \frac{4\pi}{k}\bar C^2 C_2
\end{pmatrix}
\equiv \tilde L_B + L_F\,.
\label{tildeLcircle}
\ee
So we can write $W_{1/2}-W_{1/6}$ in a power series of terms with the 
$W_{1/6}$ connection
\be
\begin{aligned}
W_{1/2}-W_{1/6}&=
\Tr_{\cal R}\cP\left[
e^{i\int_0^{2\pi} L_{1/6}d\tau}
\left(i\int_{0}^{2\pi} d\tau_1\, \tilde L(\tau_1)
-\int_{\tau_1<\tau_2}\hskip -.5cm d\tau_1d\tau_2\, \tilde L(\tau_1)\tilde L(\tau_2)
+\cdots\right)\right]\,.
\end{aligned}
\ee
As we saw in the supersymmetry analysis, terms with different numbers of 
integrals mix. It will be therefore useful to separate this sum into terms 
with different numbers of field insertions. First $L_F$, then $\tilde L_B$ and 
$L_F^2$, next $L_F\tilde L_B$ and $L_F^3$, etc.

Before finding $V$ we should choose one of the supercharges annihilating 
the $1/6$ BPS Wilson loop. We take%
\footnote{ For consistency with the analysis of \cite{Kapustin:2009kz} 
we consider here only one specific chirality.}
\be
Q
=
(Q_{12+}+i S_{12+})+
(Q_{34+}-i S_{34+})
\ee
and define
\be
\Lambda
=
i\sqrt{\frac{\pi}{2k}}\,e^{i\tau/2}
\begin{pmatrix}
0& 
C_2\\\bar C^2&0
\end{pmatrix}.
\ee
It is easy to check that
\be
Q
\,\Lambda
=L_F\,,\qquad
Q
\,L_F=
-8\cD_\tau (e^{-i\tau}\Lambda)\,,\qquad
8i e^{-i\tau}\Lambda
\Lambda=  \tilde L_B\,.
\label{Qrules}
\ee
The covariant derivative acting on $\Lambda$ 
in $Q\,L_F$ has the generalized 
connection with $\cA$ and $\hat \cA$ in $L_{1/2}$ \eqn{cA-circle}, 
but its action on $\Lambda$ is the same as 
a covariant derivative in the $L_{1/6}$ connection, since the difference between 
the two, involving $C_2\bar C^2$ and $\bar C^2 C_2$, cancels when acting 
on $\Lambda$. We can therefore integrate the total derivative inside a 
Wilson loop with either the $L_{1/2}$ or $L_{1/6}$ connection.

We now solve for $V$ in a power series. We take 
$V=\sum_{p=1}^\infty V_p$, where the term $V_p$ has 
$p$ field insertions into the Wilson loop with connection $L_{1/6}$. 
The first few are 
\begin{align}
V_1 =&\, i \Tr_{\cal R}\cP\left[
e^{i\int_0^{2\pi} L_{1/6}d\tau}\int_0^{2\pi} d\tau_1\,\Lambda(\tau_1)
\right],
\nonumber\\
V_2=&\, -\frac{1}{2} \Tr_{\cal R}\cP\left[
e^{i\int_0^{2\pi} L_{1/6}d\tau}
\int_{\tau_1<\tau_2}\hskip -.5cm d\tau_1\,d\tau_2\,
\Big(\Lambda(\tau_1)L_F(\tau_2)-L_F(\tau_1) \Lambda(\tau_2)\Big)\right],
\\
V_3 =&\, \Tr_{\cal R}\cP\bigg[
e^{i\int_0^{2\pi} L_{1/6}d\tau} 
\bigg(-\int_{\tau_1<\tau_2}\hskip -.5cm d\tau_1\,d\tau_2\,
\Big(\tilde L_B(\tau_1)\Lambda(\tau_2)+\Lambda(\tau_1)\tilde L_B(\tau_2)\Big)
\nonumber\\
&\, - i \int_{\tau_1<\tau_2<\tau_3}\hskip -1.1cm d\tau_1\,d\tau_2\,d\tau_3
\Big(\Lambda(\tau_1) L_F(\tau_2)L_F(\tau_3)
+L_F(\tau_1)\Lambda(\tau_2)L_F(\tau_3)
+
L_F(\tau_1)L_F(\tau_2)\Lambda(\tau_3)\Big)\bigg)\bigg].
\nonumber
\end{align}
Using \eqn{Qrules} it is easy to check that
\be
\begin{aligned}
Q\,V_1 =&\, i \Tr_{\cal R}\cP\left[
e^{i\int_0^{2\pi} L_{1/6}d\tau}\int_0^{2\pi} d\tau_1\,L_F(\tau_1)
\right],
\\
Q\,V_2=&\, \Tr_{\cal R}\cP\left[
e^{i\int_0^{2\pi} L_{1/6}d\tau}
\left(i\int_0^{2\pi} d\tau_1\,\tilde L_B(\tau_1)
-\int_{\tau_1<\tau_2}\hskip -.5cm d\tau_1\,d\tau_2\,L_F(\tau_1)L_F(\tau_2)
\right)\right],
\\
Q\,V_3 =&\, \Tr_{\cal R}\cP\bigg[
e^{i\int_0^{2\pi} L_{1/6}d\tau} 
\bigg(-\int_{\tau_1<\tau_2}\hskip -.5cm d\tau_1\,d\tau_2\,
\Big( \tilde L_B(\tau_1)L_F(\tau_2)+L_F(\tau_1) \tilde L_B(\tau_2)\Big)
\\
&\,\hskip4cm - i \int_{\tau_1<\tau_2<\tau_3}\hskip -1.1cm d\tau_1\,d\tau_2\,d\tau_3\,
L_F(\tau_1)L_F(\tau_2)L_F(\tau_3)\bigg)\bigg].
\end{aligned}
\ee

These indeed are the terms in the expansion of $W_{1/2}-W_{1/6}$ around 
the $L_{1/6}$ connection with one, two and three fields insertions. 
We have also checked the next term in the expansion and expect this pattern 
to extend to all orders.

This comparison with the 1/6 BPS loop allows for an alternative, immediate 
verification that the loop constructed in Section \ref{sec:loop} is indeed 1/2 BPS. 
This derivation shows that the Wilson loop is invariant under the 
single supercharge used above. Then we note that from inspecting $L_{1/2}$, 
the Wilson loop preserves an $SU(3)$ subgroup 
of the $SU(4)$ R-symmetry group of the theory. Note though that the supercharge 
is not invariant under this $SU(3)$, so acting with this symmetry we automatically 
generate more supercharges preserved by this loop. In a similar fashion 
one can generate the full supergroup with twelve supercharges preserved by the loop 
as the minimal one containing the $SU(3)$ generators and the four preserved 
by the $1/6$ BPS loop.


\subsection{Supermatrix model and supergroup Chern-Simons}

Since the 1/2 BPS loop and the 1/6 BPS one are in the same cohomology class 
with respect to $Q$, 
we can immediately conclude that the localization argument used in 
\cite{Kapustin:2009kz} for the 1/6 BPS circular loop will also apply 
unaltered to our operator.

Generalizing the matrix model derived in \cite{Kapustin:2009kz} to the 
case of $M\neq N$ gives the following expression for the partition function
\be
Z=\int \prod_{a=1}^N d\lambda_a\,
e^{ik\pi\lambda_a^2}
\prod_{\hat a=1}^M d\hat\lambda_{\hat a}\,e^{-ik\pi\hat\lambda_{\hat a}^2}
\,\frac{\prod_{a<b}\sinh^2(\pi(\lambda_a-\lambda_b))
\prod_{\hat a<\hat b}\sinh^2(\pi(\hat\lambda_{\hat a}-\hat\lambda_{\hat b}))}
{\prod_{a,\hat a}\cosh^2(\pi(\lambda_a-\hat\lambda_{\hat a}))}.
\label{ZMM}
\ee
Here $\lambda_a$ ($a=1,\ldots,N$) and $\hat\lambda_{\hat a}$ ($\hat a=1,\ldots,M$) 
are two sets of eigenvalues corresponding to the two gauge groups of the theory. 
Our 1/2 BPS Wilson loop in the fundamental representation is evaluated by inserting  
into the integral above
\be
W=\sum_{a=1}^N e^{2\pi\lambda_a}+\sum_{\hat a=1}^M e^{2\pi\hat\lambda_{\hat a}}\,.
\label{WMM}
\ee
For a general representation the insertion is (see footnote \ref{tr-footnote})
\be
W_\cR=
\Tr_\cR
\begin{pmatrix}
\diag(e^{2\pi\lambda_a})&0\\
0&\diag(e^{2\pi\hat\lambda_{\hat a}})\\
\end{pmatrix}
=\sTr_\cR
\begin{pmatrix}
\diag(e^{2\pi\lambda_a})&0\\
0&-\diag(e^{2\pi\hat\lambda_{\hat a}})\\
\end{pmatrix}
\label{general-rep}
\ee

Examining these expressions one sees that if the $\cosh$ functions 
in \eqn{ZMM} were in the numerator rather than in the denominator, this 
would be the matrix model for pure Chern-Simons theory with gauge group 
$SU(N+M)$ on a lens space $S^3/\bZ_2$, where an $SU(N)$ subgroup is 
expanded around the trivial vacuum and an $SU(M)$ subgroup around the 
other flat connection \cite{Marino:2002fk,Aganagic:2002wv,Halmagyi:2003ze}.%
\footnote{We are grateful to Marcos Mari\~no for discussions about this point.} 
Compared to the trivial saddle point with unbroken $SU(N+M)$, 
the non-trivial connection is represented by the 
shift of the eigenvalue $\lambda\to\lambda+i/2$. 
This replaces some of the $\sinh$ functions with $\cosh$ functions and also gives 
the $(-)$ factor in the lower-right block on the right-hand side of \eqn{general-rep}.

In \eqn{ZMM} the $\cosh$ functions are in the denominator rather than in the numerator. 
This arises naturally when considering instead the matrix model 
and Chern-Simons theory (and the same saddle point) for the gauge supergroup 
$SU(N|M)$. For a fuller discussion see \cite{Marino-Putrov}.


\section{Discussion}
\label{sec:discussion}

In this paper we have found the so far elusive 1/2 BPS Wilson loop operator of $\cN=6$ super Chern-Simons-matter.
We have considered both a loop supported along an infinite straight line and one supported along a circle. The former preserves separately six Poincar\'e supercharges and six conformal supercharges, whereas the latter preserves twelve  linear combinations of the two. The proof of the invariance of our operator under these supercharges is quite novel, for it requires to Taylor expand the exponential and to integrate by parts some of the variations in order to have cancellations between terms of different order. 
While the theory has only $U(N)\times U(M)$ gauge symmetry, our loop can be 
defined for arbitrary representations of the supergroup $U(N|M)$.

We have shown that this loop is related to another one, which is only $1/6$ BPS,  
by the addition of a term 
exact under a supercharge $Q$. This in turn implies that the expectation values 
of the two loops are equal and allows us to use the matrix model 
(\ref{ZMM}), derived using localization in \cite{Kapustin:2009kz}. 
Beyond this formal derivation it would be interesting to check this expression 
by an explicit perturbative computation. 
It is straightforward to do the matrix model calculation perturbatively, plugging 
(\ref{WMM}) into (\ref{ZMM}). The first few orders are
\be
\vev{W}_{MM}
=1+i\frac{\pi}{k}(N-M)
-\frac{2\pi^2}{3k^2}\left(N^2-\frac{5}{2}NM+M^2-\frac{1}{4}\right)
+{\cal O}\left(\frac{1}{k^3}\right)\,.
\label{MMprediction}
\ee 
This is a prediction for a corresponding computation to be performed directly 
in the gauge theory (with framing one) by summing Feynman diagrams \cite{inprogress}. 

The matrix model gives, in principle, an exact expression for the expectation 
value of the Wilson loop valid for  all values of $N$, $M$ and $k$. 
Unfortunately, unlike 
the 4-dimensional analog \cite{Erickson,Dru-Gross,Pestun:2007rz}, this 
matrix model is quite complicated (for similar models see {\it e.g.}  
\cite{Marino:2002fk,Aganagic:2002wv,Halmagyi:2003ze,Halmagyi:2003mm,
Marino:2009dp}). Still, 
it can be solved \cite{Marino-Putrov} and gives the correct expression for 
the expectation value of the Wilson loop at strong coupling as evaluated by 
a macroscopic fundamental string extending in $AdS_4\times \CP^3$ and 
ending along the circular loop on the boundary of $AdS_4$ 
(or by an M2 brane wrapping the orbifolded direction of $S^7/\bZ_k$)
\be
\vev{W}\simeq \exp \left(\pi\sqrt{\frac{2N}{k}}\right)
\simeq \exp \left(\pi\sqrt{\frac{N+M}{k}}\right)\,.
\ee

Note that while the matrix model is related to the supergroup Chern-Simons 
theory, it is not exactly the same. The matrix model calculates the contribution 
of a single saddle point in a perturbative expansion of the Chern-Simons theory. 
This is reminiscent of the situation for the Wilson loops on $S^2$ in $\cN=4$ 
supersymmetric Yang-Mills in four dimensions and their relation to two-dimensional 
Yang-Mills \cite{DGRT-YM2, Pestun:2009nn}. They are not given by the 
full answer in the 2-dimensional theory, but rather by a semiclassical 
expansion around the zero-instanton sector 
\cite{Witten:1992xu,Staudacher:1997kn,Bassetto:1998sr}. 
Recently an interpretation was given for the other saddle points, as the 
correlation function of Wilson and 't Hooft loops \cite{Giombi:2009ek}. 
It would be interesting to understand if there are any observables in 
$\cN=6$ Chern-Simons-matter theory which give the other saddle points 
in the perturbative expansion of the $U(N|M)$ pure Chern-Simons theory.

Another direction worth investigating is related to the construction of Wilson 
loops via the higgsing of membranes, as done in four dimensions in 
\cite{Gomis:2006sb}.  In 
\cite{Berenstein:2008dc, Rey-ABJM-wl} the coupling of the Wilson loop to the 
scalar fields was found by separating membranes and computing the mass 
of the resulting off-diagonal modes stretching between them. This indeed 
gives the scalar couplings in (\ref{MIJ}), but did not include the fermions, that,  
as we have seen, are crucial to enhance the supersymmetry of the loop 
operator. It would be therefore interesting to repeat the calculation considering 
also fermionic off-diagonal modes and to reproduce the couplings 
$\eta_I^\alpha$ and $\bar\eta^I_\alpha$ in this way.

There are other objects in this theory which are very closely related to the Wilson 
loops constructed here. These are the vortex loop operators of \cite{Dru-vortex}, 
which have a semiclassical description in the gauge theory. Along the loop the 
gauge symmetry is broken to some subgroup and different $U(1)$ factors 
have vortices. In addition, the scalar fields can have square-root branch 
cuts. So, parameterizing the transverse plane to the line by complex coordinates 
$z$ and $\bar z$, the field configuration (in one $U(1)$ factor) is
\be
A=\widehat A=-i\frac{\alpha}{2k}\left(\frac{dz}{z}-\frac{d\bar z}{\bar z}\right),\qquad 
C_1=\frac{\beta}{\sqrt{z}}\,,
\ee
with $\alpha$ and $\beta$ being two real parameters.
The vortex loops carry $k$ unit of electric flux and should be an alternative 
description for $k$ coincident Wilson loops. In fact, while one can construct the 
Wilson loop in any anti-symmetric representation, in Chern-Simons theory the dimension 
of the symmetric representation should be smaller than $k$. We expect the 
vortex loops to take over as the description of the object carrying $k$ units of 
electric flux. In the M-theory picture the $1/2$ BPS Wilson loop is an M2-brane 
wrapping the orbifolded direction of $S^7/\bZ_k$. The $k$-th symmetric 
loop is the brane wrapping the circle in the covering space. This brane then 
develops extra allowed deformations, including opening up in $AdS_4$, which 
corresponds to the $\beta$ parameter above, and rotating on $S^7$, thus leading to the 
$1/3$ BPS vortex of \cite{Dru-vortex}.

Still, a fuller classification of all 1-dimensional defects in this theory is in order. 
In $\cN=4$ super Yang-Mills in four dimensions the classification of 
Wilson and 't~Hooft loops gives rise to a rich structure of objects in the dual 
string theory, including probe branes wrapping various cycles (see {\it e.g.} 
\cite{Dru-Fiol-giant,Yamaguchi:2006tq,Gomis:2006sb,Gomis:2006im}) and 
fully backreacted geometries, the so-called ``bubbling'' solutions (see {\it e.g.} 
\cite{Yamaguchi:2006te,Lunin:2006xr,D'Hoker:2007fq,Okuda:2008px}). 
So far this classification has only been partially undertaken in the M-theory dual of $\cN=6$ 
super Chern-Simons-matter. We plan to complete this in a future publication 
\cite{inprogress}.

Apart for the theory with $\cN=6$ supersymmetry, there are closely related 
3-dimensional Chern-Simons-matter theories with $\cN=4$ and $\cN=5$ 
supersymmetry \cite{Gaiotto:2008sd,Hosomichi:2008jd,Hosomichi:2008jb}. 
These theories are based on more complicated quivers and have a richer 
structure of allowed gauge groups and matter representations. We expect 
that constructions similar to ours will give the $1/2$ BPS loops of these 
theories. These should probably be related to Wilson loop observables 
in the topological theories discussed in \cite{Kapustin:2009cd}.

It would be also of some importance to find other loop operators of 
${\cal N}=6$ super Chern-Simons-matter preserving reduced amounts of 
supersymmetry, following the spirit of \cite{Zarembo:2002an,DGRT-big}. 
In particular, note that the couplings satisfy the relations
$\bar\eta_\alpha^I\eta^\beta_I=i(1+\dot x^\mu\gamma_\mu)_\alpha^{\ \beta}$
and $M^I_J=\delta^I_J+i\eta^\alpha_J\bar\eta_\alpha^I$. Such relations may 
play a role in a more comprehensive analysis as do the pure spinors 
in the treatment of \cite{Dymarsky:2009si}.


\subsection*{Acknowledgements}

This project grew out of discussions with Georgios Michalogiorgakis and
Vasilis Niarchos to whom we are extremely grateful for the initial collaboration.
We would also like to thank David Berenstein, Giulio Bonelli, 
Jaume Gomis, Johannes Henn, 
Juan Maldacena, Jan Plefka, and especially Marcos Mari\~no 
for interesting discussions. We also thank Marcos Mari\~no and Pavel Putrov 
for sharing their results \cite{Marino-Putrov} prior to publication. 
We thank the KITP for hospitality during the ``Fundamental Aspects
of Superstring Theory'' workshop. N.D. would also like to thank the hospitality of
the \'Ecole Polytechnique in Paris, the Benasque Science Center, 
and the hospitality and financial support of
Perimeter Insititute. D.T. is grateful to the
Theory Group at CERN, the LPTHE in Paris, the Niels Bohr Institute, and the 7th Simons Workshop in Stony Brook for the very nice
hospitality during the completion of this work. N.D. and D.T. are partly
supported by the NSF grant PHY-05-51164 and D.T. also by the Department of
Energy under Contract DE-FG02-91ER40618.


\appendix

\section{Notation and conventions}
\label{app-SUSY}

For the supersymmetry analysis of the loop, we consider the theory in $\mathbb{R}^{1,2}$ with metric $g_{\mu\nu}=\diag(-1,1,1)$ and space-time indices $\mu,\nu,\ldots=0,1,2$. The spinor indices are
denoted with lower case letters from the beginning of the Greek alphabet,
$\alpha,\beta,\ldots=+,-$. Spinor indices are raised and lowered according
to the following rules
\be
\begin{aligned}
\psi^\alpha=\varepsilon^{\alpha\beta}\psi_\beta\,,\qquad
\psi_\alpha=\varepsilon_{\alpha\beta}\psi^\beta\,,\qquad
\varepsilon^{+-}=-\varepsilon_{+-}=1\,,
\end{aligned}
\ee
and we choose for our basis of gamma matrices
\be
(\gamma^\mu)_\alpha^{\ \beta}=\{-i\sigma^3,\sigma^1,\sigma^2\}\,,
\label{gamma-basis}
\ee
obeying the relation $\gamma^\mu\gamma^\nu=g^{\mu\nu}+\varepsilon^{\mu\nu\rho}
\gamma_\rho$ (with $\varepsilon^{012}=1$). Lowering the upper index,
these matrices become symmetric
\be
(\gamma^\mu)_{\alpha\beta}=\{-i\sigma^1,-\sigma^3,i 1\}\,.
\ee
When not written explicitly, the spinor indices are
contracted as
\be
\begin{gathered}
 \theta\psi\equiv\theta^\alpha \psi_\alpha=-\theta_\alpha\psi^\alpha
=\psi^\alpha\theta_\alpha=\psi\theta\,,
\\
\theta\gamma^\mu \psi \equiv \theta^\alpha
(\gamma^\mu)_{\alpha}^{\ \beta}\psi_\beta=
-\psi\gamma^\mu\theta\,,
\end{gathered}
\ee
where $\theta$ and $\psi$ are arbitrary spinors.

Regarding the gauge index structure, if we denote by $a$, $\hat a$ the
gauge indices in the fundamental of the first and the second
gauge group, respectively, we have
\be
\label{prelaa-app}
(C_I)_a^{~\hat a}\,,\qquad (\bar C^I)_{\hat a}^{~a}\,,
\qquad 
(\psi_I)_{\hat a}^{~a}\,,\qquad (\bar\psi^I)_a^{~\hat a}\,.
\ee

Under supersymmetry the fields transform as (for clarity we indicate here all the spinor indices explicitly)
\begin{align}
\delta A_\mu&=\frac{4\pi i}{k} \bar \theta^{IJ\alpha} (\gamma_\mu)_\alpha^{\ \beta}
\left(C_I \psi_{J\beta}+\frac{1}{2}\varepsilon_{IJKL}\bar \psi^K_\beta \bar C^L\right)\,,
\cr
\delta  \widehat A_\mu&=\frac{4\pi i}{k} \bar \theta^{IJ\alpha} (\gamma_\mu)_\alpha^{\ \beta}
\left(\psi_{J\beta} C_I+\frac{1}{2}\varepsilon_{IJKL}\bar C^L\bar \psi^K_\beta\right)\,,
\cr
\delta C_K&= \bar \theta^{IJ\alpha} \varepsilon_{IJKL} \bar \psi^L_\alpha\,,
\\
\delta \bar C^K&=2\bar \theta^{KL\alpha}\psi_{L\alpha}\,,
\cr
\delta\psi_K^\beta&=-i\bar \theta^{IJ\alpha}\varepsilon_{IJKL}
(\gamma^\mu)_{\alpha}^{\ \beta} D_\mu\bar C^L\cr & \hskip 2cm 
+\frac{2\pi i}{k}\bar \theta^{IJ\beta}\varepsilon_{IJKL} \big(\bar C^LC_P\bar C^P-\bar C^PC_P\bar C^L\big)
+\frac{4\pi i}{k}\bar \theta^{IJ\beta}\varepsilon_{IJML}\bar C^MC_K\bar C^L,
\cr
\delta\bar\psi^{K}_\beta&=-2i\bar\theta^{KL\alpha}
(\gamma^\mu)_{\alpha\beta} D_\mu C_L-\frac{4\pi i}{k}\bar\theta^{KL}_\beta(C_L\bar C^M C_M-C_M\bar C^M C_L)
-\frac{8\pi i}{k} \bar\theta^{IJ}_\beta C_I\bar C^K C_J\,,
\nonumber
\end{align}
where we have written the transformations only in terms of the parameters $\bar\theta$ and not $\theta$, by using the following relation
\be
\theta_{IJ}=\frac{1}{2} \varepsilon_{IJKL} \bar\theta^{KL}\,.
\ee
The supersymmetry parameters are antisymmetric, $\bar\theta^{IJ}=-\bar\theta^{JI}$, and obey the reality condition 
\be
\bar\theta^{IJ}= (\theta_{IJ})^*\, . 
\label{real}
\ee

\bibliography{refs}
\end{document}